\documentclass[prl,twocolumn,showpacs,superscriptaddress]{revtex4-1}
\usepackage{graphicx,amssymb,amsmath,color}
\begin{document}
\definecolor{darkgreen}{rgb}{0,0.5,0}
\newcommand{\rim}[1]{\textcolor{green}{#1}}

\title{Luttinger liquid with complex forward scattering: robustness and Berry phase}

\author{Bal\'azs D\'ora}
\email{dora@eik.bme.hu}
\affiliation{Department of Physics and BME-MTA Exotic  Quantum  Phases Research Group, Budapest University of Technology and
  Economics, 1521 Budapest, Hungary}
\author{Roderich Moessner}
\affiliation{Max-Planck-Institut f\"ur Physik komplexer Systeme, 01187 Dresden, Germany}

\date{\today}

\begin{abstract}
Luttinger liquids (LLs) are one dimensional systems with well-understood  instabilities due to umklapp or backscattering.
We study a generalization of the Luttinger model, which incorporates a time reversal symmetry breaking 
interaction producing a complex forward scattering 
amplitude ($g_2$ process). The resulting
low energy state is still a LL, and belongs to the
family of interacting Schulz-Shastry models. Remarkably, it  becomes increasingly robust against 
additional perturbations -- for purely imaginary $g_2$, both umklapp and local backscattering
are  always irrelevant. Changing the phase of the interaction generates a non-trivial 
Berry phase, with a universal geometric phase difference between ground and  a one boson excited 
state depending only on the LL parameter.
\end{abstract}

\pacs{71.10.Pm,05.30.Fk,05.70.Ln,67.85.-d}

\maketitle

\paragraph{Introduction.}
Landau Fermi liquid theory enjoys great success in describing  three dimensional metals. 
In particular, through the quasiparticle concept, it  
explains why interacting fermions behave similarly to a Fermi gas.

However, upon reducing  spatial dimensionality, instabilities of the gas become more pronounced
and Fermi liquid theory loses its applicability. This is most striking  in one dimension,
where the conventional quasiparticles break down. They are often replaced by 
collective bosonic excitations described by Luttinger liquid (LL) theory.
LL physics is not restricted to  condensed matter physics\cite{giamarchi,nersesyan} but 
plays a role  whenever confinement to one 
spatial dimension is strong, e.g.\ in field theories of high energy physics (e.g. massless Thirring model\cite{thirring}), cold atoms\cite{cazalillarmp}, 
or the study of black  holes\cite{LLblackhole}.

These collective bosonic excitations can further become unstable, in the presence of 
umklapp or backscattering due to interactions or impurities, as the LL undergoes a 
phase transition to a different state: 
stabilizing the LL low energy modes represent a difficult task.

To harness correlations and entanglement of LL -- e.g.\ in a putative quantum computer \cite{nielsen} -- 
it is desirable for them to be robust to additional perturbations. Besides isolating a LL carefully from 
its environment,
additional stability can either be achieved by prohibiting backscattering through carefully cancelling  
dangerous terms in the LL Hamiltonian; or by {\it adding} extra terms, which act to suppress the effects of
backscattering while preserving the low energy properties of the LL.
Here we follow this latter route by designing a Luttinger liquid with extremely robust collective low energy modes.
This is achieved by introducing an interaction breaking time reversal symmetry, 
which changes from attractive to repulsive depending on whether  
two particles move towards or away from each other.


We first introduce the Hamiltonian of this generalized LL. We then  
analyze its properties, demonstrating its immunity with respect 
to umklapp and backscattering as a function of the phase of its complex coupling constant $g_2$. We 
investigate the universality of its geometric phase and finally discuss ingredients for its possible realization.

\paragraph{The model.}
The low energy physics of many types (fermions, bosons, spins) of 
one dimensional interacting particles  is described by the Luttinger model. 
We study its generalization to include a \emph{complex} $g_2$  process:
\begin{gather}
H=\sum_{q\neq 0} \omega(q)b_q^\dagger b_q+\frac{g_q}{2}b_qb_{-q}+\frac{g_q^*}{2}b_q^+b_{-q}^+,
\label{hamilton}
\end{gather}
with $\omega(q)=v|q|$, $v$ the  "sound velocity",  
$b_q^\dagger$ the creation operator of a bosonic density wave and 
$g_q=g_2|q|\exp(i\varphi)$ 
parametrising the interparticle interaction.
We have neglected velocity renormalization\cite{giamarchi,solyom}. 
The conventional case, discussed thoroughly in 
Refs. \cite{giamarchi,nersesyan,solyom} corresponds to $\varphi=0$,  where forward 
scattering (small momentum transfer compared to the Fermi momentum, $k_F$) interactions are included.
The whole parameter space is covered by  
arbitrary real values of $g_2$ (describing both attractive and repulsive interactions), and restricting the phase to  
$|\varphi|\le\pi/2$. Other values of the phase are accounted for by changing the sign of 
$g_2$.

Eq.~\eqref{hamilton} resembles the Bogoliubov Hamiltonian of a Bose-Einstein condensate, with $\varphi$ 
playing the role of the condensate phase. This phase drops out from 
most physical observables
and correlation functions,  being visible primarily in interference probes. 
We show that for LLs, the innocent-looking phase variable
has profound effect on both correlators and the stability of the LL.

The origin of phase $\varphi$  is readily illustrated
for interacting spinless fermions.
The conventional real part of the $g_2$ process 
is due to a short-range forward scattering density-density interaction \cite{giamarchi}, 
$H_{re}=g^\prime\int dx \rho_R(x)\rho_L(x)$, where $\rho_{R}(x)=:R^+(x)R(x):$ and 
$\rho_{R}(x)=:L^+(x)L(x):$ are  normal ordered densities, and $R(x)/L(x)$ annihilates a
right/left moving particle at point $x$, respectively.

The novel aspect of our work, the imaginary part of the interaction,
can be generated from a long range interaction between the above densities
\begin{gather}
H_{im}=-2g^{\prime\prime} \int dx\int dy \frac{\rho_R(x)\rho_L(y)}{x-y}.
\label{hprime}
\end{gather}
This breaks  time reversal invariance and has the same scaling dimension (also marginal) 
as the above $g^\prime$ process. Therefore
they should be considered on equal footing.

This interaction is repulsive/attractive when two particles move away from/towards each other (or the opposite for both,
depending on the sign of $g^{\prime\prime}$). In other words, the repulsive or attractive nature of the interaction 
depends on the relative motion of the interacting pair.
The self-energy correction of $H_{im}$ from diagrammatics give the same 
LL behaviour as the $H_{re}$. However,  singular vertex
corrections are absent, in sharp contrast to the conventional case, see below.

Bosonizing $H_{im}$ following Ref.~\cite{giamarchi} yields imaginary $g_2$:
\begin{gather}
H_{im}=ig^{\prime\prime}\sum_{q\neq 0} \frac{|q|}{2}\left(b_q^+b_{-q}^+-b_qb_{-q}\right)\ .
\end{gather}
Keeping both $H_{re}$ and $H_{im}$ gives Eq.~\eqref{hamilton} with 
$g_2\cos\varphi=g^\prime$ and $g_2\sin\varphi=g^{\prime\prime}$.

The phase can be `gauged away' from Eq.~\eqref{hamilton} by a unitary 
transformation leaving the eigenenergies and
 the renormalized velocity $\sqrt{v^2-g_2^2}$ unchanged, with LL parameter $K=\sqrt{(v-g_2)/(v+g_2)}$.
The sign change of $g_2$ amount to $K\rightarrow 1/K$ change.
Eq.~\eqref{hamilton} is diagonalized by a Bogoliubov rotation to bosonic operators $c_q$: 
\begin{gather}
b_q=\frac{K+1}{2\sqrt K}c_q+\exp(i\varphi)\frac{K-1}{2\sqrt K}c^+_{-q},
\label{bogtrafo}
\end{gather}
yielding $H=\sum_{q\neq 0} \sqrt{v^2-g_2^2}|q|c^+_qc_q$. The essential change with respect to 
the  conventional case is the phase factor in Eq.~\eqref{bogtrafo}.

\paragraph{Correlation functions.}
While the spectrum does not change, correlation functions do. Let us consider
an underlying spinless fermion field which decomposes 
decomposes into  right- and left-going parts, $\Psi(x)= 
e^{ik_F x}R(x) +e^{-ik_F x}L(x)$. To evaluate 
the right-going Green function 
\begin{gather}
G_R(x,t)\equiv \langle R^+(x,t)R(0,t)\rangle\ ,
\end{gather}
$R(x)$ is expressed in terms of the real space version of the $b$ operators
via \cite{giamarchi} $R(x)=\frac{1}{\sqrt{2\pi\alpha}}\exp\left[i(\phi(x)-\theta(x))\right]$ 
with $[\phi(x),\theta(y)]=i\frac \pi 2\textmd{sign}(y-x)$ the commutation relation of the dual fields. These are related to the $b$ bosons as $\phi(x)=\sum_{q}\sqrt{2\pi/
  |q|L}e^{iqx-\alpha |q|/2}b_q+h.c.$ and $\theta(x)=\sum_{q}\textmd{sign}{q}\sqrt{2\pi/ |q|L}e^{iqx-\alpha |q|/2}b_q+h.c.$ and $\alpha$ is the short distance cutoff.
Standard analysis~\cite{giamarchi,nersesyan} yields
$G_R(x,t)\sim |x|^{-\eta_R}$
with
\begin{gather}
\eta_R=\frac{K+K^{-1}}{2}.
\label{exp_boson}
\end{gather}
independent of $\varphi$. The  left-movers behave identically.
Nonetheless, for the correlators of the dual fields,
we find
\begin{gather}
\left\langle \left( \phi(x,t)-\phi(0,t)\right)^2\right\rangle=\eta_\phi\ln(x),
\label{phiphi}
\end{gather}
\begin{gather}
\eta_\phi=K\cos^2\left(\varphi/2\right)+({1}/{K})\sin^2\left({\varphi}/{2}\right)\ .
\end{gather}
Conventional duality of LLs for $\langle \left( \theta(x,t)-\theta(0,t)\right)^2\rangle=\eta_\theta\ln(x)$, using the $K\rightarrow 1/K$ change for $\eta_\theta$, 
gives
\begin{gather}
\eta_\theta=(1/K)\cos^2\left(\varphi/2\right)+K\sin^2\left({\varphi}/{2}\right).
\end{gather}

The dominant instabilities for spinless fermions are expected in the $2k_F$ charge 
density wave or in the superconducting channel, with order parameters
$O_{CDW}=R^+(x)L(x)\sim \exp(-i2\phi(x))$ and $O_{SC}=R(x)L(x)\sim \exp(-2i\theta(x))$, respectively.
These decay with respective exponents $-2\eta_\phi$ and $-2\eta_\theta$.
For $\cos(\varphi)>0$, the conventional conclusions stand\cite{giamarchi}: 
for $0<K<1$ (attractive interactions), the SC instability dominates, while for repulsive $K>1$,
density wave ordering is favoured. For $\cos(\varphi)<0$, however, the reverse is the case.

In between, for $\varphi=\pm\pi/2$, the interaction term is purely imaginary, 
and the exponents are equal,
$2\eta_\phi=2\eta_\theta=K+K^{-1}>2$, exactly
twice $\eta_R$ of the single particle Green function, Eq.~\eqref{exp_boson}.
This indicates the absence of singular vertex corrections
due to the peculiar attractive-repulsive behaviour of the interaction,  Eq.~\eqref{hprime}.
The correlation functions thus decay with an exponent  bigger than 2, implying a \emph{faster} decay to zero than in the non-interacting case in both particle-hole 
and Cooper channels. As a result, these
instabilities are excluded!

\paragraph{Umklapp and backscattering.}
To assess the stability of this LL, one has to consider 
further scattering processes which could induce a gap.
One is umklapp scattering, arising at half filling for spinless particles:
\begin{gather}
H_u=\frac{g_u}{(2\pi\alpha)^2}\int dx \cos(4\phi(x)).
\end{gather}
From Eq.~\eqref{phiphi}, $\langle \cos(4\phi(x))\cos(4\phi(y))\rangle \sim |x-y|^{8\eta_\phi}$, 
yielding scaling dimension 
 $4\eta_\phi$\cite{nersesyan}.
In 1+1=2 space-time dimensions of the model, 
the umklapp term thus is relevant for $4\eta_\phi<2$.
This reproduces the result $K<1/2$ for a conventional LL ($\varphi=0$) \cite{giamarchi}.
For the general case, the perturbation can only be relevant ($\eta_\phi<1/2$) for $|\varphi|<\pi/6$, otherwise
it is irrelevant regardless of the value of $K$. 
For spinful fermions, e.g.\ 
in the charge sector of the Hubbard model, 
umklapp scattering gives rise to a term with $\cos(\sqrt 8 \phi(x))$, relevant
for $\eta_\phi<1$. 

\begin{figure}[h!]
\centering
{\includegraphics[width=6.cm]{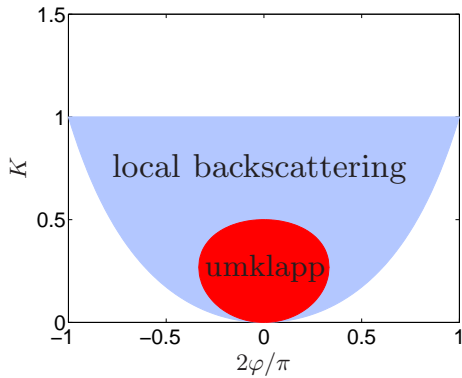}}
\caption{(Color online)  
The $K$ -- $\varphi$ phase diagram for spinless fermions, the colored regions denote the relevance of the 
umklapp processes (red) and local backscattering (blue).
The latter region also denotes the relevance of the umklapp scattering for spinful fermions in e.g. the Hubbard model.
For cosine terms containing $\theta$, $K$ should be changed to $1/K$.}
\label{phasediag}
\end{figure}

The $K$ -- $\varphi$ phase diagram for these two cases is plotted in Fig.~\ref{phasediag}.
For purely imaginary interaction, $\varphi=\pm\pi/2$, these sine-Gordon-like terms are all irrelevant and 
{\it the LL is stable for any $K$}, even at commensurate
filling. This is due to the lack of vertex corrections from Eq.~\eqref{hprime}.

Local perturbations such as potential scattering can modify the transport properties of a LL. 
Depending on the LL parameter, it can flow to strong coupling and effectively cut the LL
in two, or to zero and disappear from the low energy dynamics.
For the spinless case, local backscattering takes the form
\begin{gather}
H_{bs}=\frac{g_{bs}}{\pi\alpha}\cos(2\phi(0)).
\end{gather}
Since this interaction is spatially localized, 
only its temporal fluctuation need to be considered \cite{nersesyan}.
It is relevant for $\eta_\phi<1$ (as for the conventional LL \cite{giamarchi,nersesyan}),
just as umklapp scattering in the spinful case, 
Fig. \ref{phasediag}.

\paragraph{Geometric phase related to $\varphi$.}
We now turn to the properties treating the phase $\varphi$ as a tunable degree of freedom.
A non-interacting fermionic version of Eq. \eqref{hamilton} was considered in Refs. \cite{carollo,hamma}, 
and its
Berry phase signaled quantum phase transitions.
Our model, on the other hand, contains interacting fermionic degrees of freedom and is always at criticality with
continuously varying critical exponents so that 
its  geometric phase
can be sensitive to both criticality and interactions.

In the Luttinger model, a 
given momentum $q$ mode only interacts with its $-q$ partner. It therefore is suffices to analyze a 
single $(q,-q)$ pair in Eq.~\eqref{hamilton}. This is identical to 
a quantum parametric amplifier\cite{gerry} with a time dependent $\varphi$.
The $b^+_qb_q-b^+_{-q}b_{-q}$ being an unbounded constant of motion,  
an appropriate $\varphi(t)$ can enhance the boson occupation numbers and yield squeezing\cite{carmichael}.
Here we investigate the effect of an adiabatic change of this phase in Eq. \eqref{hamilton} from 0 to $2\pi$.
The calculation of the geometric phase related to this cycle follows similar steps in the Dicke model\cite{chenpra}.
The phase is induced in Eq. \eqref{hamilton} by the unitary transformation
\begin{gather}
U_R=\exp\left(-i\frac{\varphi}{2}\sum_{q\neq 0}b_q^\dagger b_q\right),
\end{gather}
with concomitant ground state wavefunction transformation
$|\Psi(\varphi)\rangle=U_R|\Psi(0)\rangle$. 
Thence, 
\begin{gather}
\gamma_g=i\int\limits_0^{2\pi}d\varphi\left\langle \Psi(\varphi)\right|\partial_\varphi \left|\Psi(\varphi)\right\rangle=\pi\sum_{q\neq 0} \langle b_q^\dagger b_q\rangle.
\end{gather}
For a given mode, the geometric phase is independent of momentum, from Eq. \eqref{bogtrafo}:
\begin{gather}
\tilde\gamma=\frac{\pi}{4}\left(K+\frac 1K-2\right),
\label{berryq}
\end{gather}
The overall geometric phase $\gamma_g=L\tilde\gamma/\pi\alpha$ 
is not universal as it depends on the high energy cutoff $\alpha$.

The elementary excitations in a LL lose their original fermionic character and statistically transmute into 
bosons \cite{nersesyan}.
It is thus interesting to consider the relative geometric phase between a nonzero momentum excited state with one boson and the ground states\cite{carollo}, given by the
difference of the respective Berry phases. 
A one-boson excited state with momentum $k$ 
is $c^+_k|\Psi(\varphi)\rangle$.
Then
\begin{gather}
\gamma_{eg}=\gamma_e-\gamma_g=i\int\limits_0^{2\pi}d\varphi\left\langle \Psi(\varphi)\right|c_k \partial_\varphi c^+_k\left|\Psi(\varphi)\right\rangle=-\tilde\gamma
\nonumber
\end{gather}
is universal and depends only on the LL parameter $K$.
This is not topological but geometric in nature and can be tuned arbitrarily  by changing the strength of the interaction.
The Berry phase difference between any two one-boson  excited states vanishes.


$\gamma_g$  also follows from the fact that the terms in the Hamiltonian
for each $(q,-q)$ pair
 are the generators\cite{dorapdf} of the SU(1,1) Lie algebra
\footnote{In particular, $J_z(q)=(b^+_qb_{q}+b_{-q}b^+_{-q})/{2}$, $J_+(q)=b^+_qb^+_{-q}$, $J_-(q)=b_qb_{-q}$
 are the generators of a
SU(1,1) Lie algebra with commutation relation $[J_+(q),J_-(q)]=-2J_z(q)$, $[J_z(q),J_\pm(q)]=\pm J_\pm(q)$.}.
The geometric
phase for general time dependent parameters for the SU(1,1) case were calculated in Ref. \cite{hongyi,gerry}, which also gives  Eq. \eqref{berryq}.
Due to the bosonic algebra in LLs, 
the geometric phase is related to the surface area on the unit hyperboloid (and not the unit sphere, as in the case 
for fermions) enclosed by the  adiabatic path traced out by $U_R(t)$.

\paragraph{Connection to interacting gauge theory.}
Gauge fields and their role in statistical transmutation in various dimensions 
are a fascinating subject of research\cite{rabello}.
A gauge potential enters into the Hamiltonian of a LL, written in terms of the dual fields as
\begin{gather}
H=\frac{\tilde v}{2\pi}\int dx \frac{1}{\tilde K}\left[\partial_x\phi(x)\right]^2+\tilde K\left[\Pi(x)-A(x)\right]^2,
\label{ham0}
\end{gather}
where $\Pi(x)=\partial_x\theta(x)/\pi$,
$\tilde K$ and  $\tilde v$ are LL parameter and renormalized velocity, respectively.
For a particular long range density dependent gauge potential of strength $\nu$,
\begin{gather}
A(y)=\nu \int dx \frac{\partial_x\phi(x)}{x-y}
\label{vectpot}
\end{gather}
the second term in Eq. \eqref{ham0} generates $H^\prime$ in Eq. \eqref{hprime},
and $\partial_x\phi(x)=-\pi[\rho_R(x)+\rho_L(x)]$ is the density operator for long wavelength excitations.

Eq.~\eqref{ham0} belongs to the family of Schulz-Shastry  models\cite{schulz,phamepl}.
$A(y)$ appears in Ref.~\cite{aglietti}, and is also common in the 
factorization of the Calogero-Sutherland model\cite{shastry}.
It represents the Hilbert transform of the charge density, acting 
over the wavefunction as the current operator\cite{fradkin}.
$A(y)$ can be unitarily transformed away \cite{schulz} from Eq. \eqref{ham0}
to yield a conventional LL form,
\begin{gather}
\exp(iS)\Pi(y)\exp(-iS)=\Pi(y)+\nu  \int dx \frac{\partial_x\phi(x)}{x-y}
\label{unitary}
\end{gather}
with $S=\nu \int dx_1\int dx_2 \phi(x_1)\phi(x_2)/(x_1-x_2)^2=\nu \int dx_1\int dx_2 \partial_x\phi(x_1)\partial_x\phi(x_2)\ln|x_1-x_2|$. Note that the latter form of $S$ describes a logarithmic long range interaction between the 
densities.

Eq. \eqref{unitary} can  be regarded as a generalized Jordan-Wigner transformation
\cite{schulz,penc,phamepl}, 
$\theta(y)\rightarrow \theta(y)-\pi\nu \int dx \ln|x-y| \partial_x\phi(x)$. 
Unlike a local vector potential $A_l(x)=\nu \partial_x\phi(x)$, which yields
anyons in 1D, the present transformation does not change the statistics of the original fields.

\paragraph{Relation to experiments.}
Here we list some ingredients for obtaining a LL with complex $g_2$, and give
pointers for how they may perhaps be realized. 
Since the gauge potential can be transformed away, the thermodynamics 
of Eqs.~\eqref{hamilton},~\eqref{ham0} is identical to that of a conventional LL. 
However, the unitary transformation has to be reinserted to compute 
correlators. 
This suggests one simulation strategy for the gauge transformed system in a quench experiment, 
since the unitary transformation can formally be identified with a time evolution operator: 
by preparing a one dimensional system in its LL ground state\cite{cazalillarmp},
adding a spatially long range logarithmic interaction, 
for a duration proportional to the strength of the gauge field $\nu$,
mimics 
$S$ and the unitary transformation in Eq.~\eqref{unitary}.
A logarithmic interaction can e.g. be approximated  
in dielectric films endowed with permeability much higher than the surrounding medium\cite{keldysh}.
The equal time correlation functions
evaluated with the quenched wavefunction
would then be identical to those from  Eq. \eqref{ham0}.

Alternatively, a density dependent, possibly long range gauge potential, Eq.~\eqref{vectpot},
can in principle be created  in a cold atomic setting using ideas similar to Refs. 
\cite{keilmann,edmonds,greschner,goldman}. More directly, a 
repulsive or attractive interaction for particles moving  away from or towards each other
is reminiscent of the physics of Doppler cooling: the frequency of a photon emitted 
by one atom as seen by the other 
depends on the direction of their velocity difference  though the Doppler effect, so that 
frequency-dependent absorption could generate the above effect.

Finally, coupling to chiral fields of fermionic\cite{yao}, 
bosonic or other origin, also provides the desired interaction $H^\prime$. On integrating out the
chiral field, $\gamma$, which couples as $[\rho_R(x)+i\rho_L(x)]\gamma(x)+\textmd{h.c.}$,
a long range interaction between the densities arises, mediated
by the field propagator $\langle \gamma(x)\gamma^+(y)\rangle\sim i/(x-y)$.

While the experimental obstacles appear formidable, it seems clear that there is no fundamental barrier
to realizing our generalized LL. Given its interesting properties, in particular
its unusual stability, we hope that it will prove to be worth the requisite effort. 


\begin{acknowledgments}

This research was  
supported by the Hungarian 
Scientific  Research Funds Nos. K101244, K105149, K108676, by the 
Bolyai Program of the HAS, and by Helmholtz VI 521 "New States of Matter
and Their Excitations". 
\end{acknowledgments}

\bibliographystyle{apsrev}
\bibliography{wboson}

\end{document}